\documentclass{article}
\usepackage{amsmath, amsthm, tikz, graphicx, cite, color, braket, subcaption, setspace} 
\usepackage[margin = 2cm]{geometry}
\usepackage[colorlinks = true]{hyperref}

\newtheorem{example}{Example}

\newtheorem{procedure}{Procedure}

\DeclareMathOperator{\diag}{diag}

\title{Quantum state transfer on a scalable network under unital and non-unital noise}

\author{
Monika Rani$^1$\thanks{Email: \texttt{rani.2@iitj.ac.in}}, Subhashish Banerjee$^1$, Nikhil Swami$^2$, Supriyo Dutta$^2$\thanks{Email: \texttt{dosupriyo@gmail.com}} \\
\small{$^1$Department of Physics, Indian Institute of Technology Jodhpur, Jodhpur, Rajasthan, India-342037.}\\
\small{$^2$Department of Mathematics, National Institute of Technology Agartala, Jirania, West Tripura, India - 799046.}
}

\date{} 
\begin{document}

\maketitle

\begin{abstract}
    We investigate quantum state transfer on a class of bipartite graphs, namely the butterfly graphs, within the framework of discrete-time quantum walks. These graphs facilitate the construction of scalable quantum networks that enable communication between a sender and a receiver via perfect state transfer. Our analysis demonstrates that state transfer occurs across different butterfly graphs, thereby extending the known families of networks that support high-fidelity quantum state transfer. 
    In addition to the ideal noiseless dynamics, we further investigate the robustness of quantum state transfer in the presence of non-Markovian environmental noise, specifically, random telegraph noise, modified Ornstein–Uhlenbeck noise, which are examples of unital noise and non-Markovian amplitude damping noise, non-unital noise. These noise models capture different types of system–environment interactions and memory effects that influence the coherence of the quantum walk. These findings contribute to the theoretical understanding of how butterfly graph constructions influence quantum transport phenomena.
\end{abstract}

\tableofcontents

\section{Introduction}
    
    In quantum communication, the concept of quantum state transfer \cite{bose2003quantum, christandl2004perfect, matsukevich2004quantum, cirac1997quantum} is fundamental. It enables the transmission of an arbitrary quantum state from one location (sender) to another (receiver) in a quantum network. This process applies to quantum technologies, including long-distance quantum communication \cite{duan2001long, muralidharan2016optimal, aspelmeyer2003long}, distributed quantum computing \cite{beals2013efficient, yimsiriwattana2004distributed, cacciapuoti2019quantum}, and quantum sensor networks \cite{eldredge2018optimal}. Hence, it is essential to perform state transfer efficiently in quantum networks to construct scalable and robust quantum systems. Quantum state transfer has been investigated across various experimental and theoretical platforms, such as spin chains, photonic systems, and quantum walks on graphs. In quantum information and computation, quantum walks \cite{aharonov1993quantum, venegas2012quantum} provide a mathematical framework for establishing a model of universal quantum computation \cite{childs2009universal}. The Discrete-Time Quantum Walks (DTQW) \cite{lovett2010universal, kurzynski2011discrete} are a class of quantum walks, which are especially interesting due to their step-wise evolution following a unitary evolution operator composed of a coin operator and a shift operator. In this article, we investigate quantum state transfer in a framework of DTQW. 
     
    In classical random walks on graphs \cite{xia2019random, childs2002example}, the state of a walker, moving from one vertex to another by randomly selecting a neighboring vertex, is updated by a transition matrix. In contrast, in a DTQW, this random choice is replaced by applying a quantum coin operator $C$, which alters the internal state of the walker. A shift operator $S$ then moves the walker to adjacent vertices based on this internal state. The overall evolution is governed by the repeated application of the unitary operator $U = S  \times C$ \cite{cao2019quantum}.
    
    Several investigations have established the usefulness of DTQWs as a framework for quantum state transfer. In \cite{vstefavnak2016perfect}, the notion of almost perfect state transfer under DTQW was introduced. Most existing research focuses on quantum state transfer over specific graph families.  Thus, for example, \cite{dutta2026perfect} demonstrates that perfect state transfer can be achieved on path and cycle graphs even in the presence of Markovian noise. To investigate quantum state transfer under DTQW, more complex graph topologies have also been studied, including star graphs, complete bipartite graphs \cite{vstefavnak2017perfect, santos2022quantum} and structured networks such as the butterfly graphs \cite{cao2019quantum}. The feasibility of  perfect state transfer over arbitrary distances feasible by a reasonable choice of coin operators was shown in \cite{yalccinkaya2015qubit}.
    
    In this work, we investigate state transfer under DTQW on a growing family of graphs, namely the butterfly graphs. These graphs are constructed systematically from path graphs. The choice of butterfly graphs is motivated by a number of significant advantages for building a quantum network. All butterfly graphs generated from $P_2$, the path graph of two vertices, are planar, which allows them to be embedded on a two-dimensional surface. This is a key requirement for designing physical quantum hardware architectures. Secondly, the butterfly graphs generated from $P_n$, the path graph with $n$ vertices, have a diameter $(n + 1)$. This means that the distance between any two vertices is at most $(n + 1)$. This property ensures rapid and efficient state transfer over the network. Moreover, the family of butterfly graphs is scalable, that is, we can increase the system size by adding more elements to the network. Therefore, it satisfies the scalability criterion of Divincenzo \cite{divincenzo2000physical}. In addition, each member of the family of butterfly graphs supports high-fidelity state transfer. In combination, the butterfly graphs ensure scalability and structural regularity, which makes them suitable for quantum networks that need to grow in size without compromising transfer efficiency. We provide analytical and numerical evidence that all members of this graph family support efficient state transfer between any pair of vertices. We also demonstrate that this framework enables the construction of large quantum networks with predictable and controllable transfer properties.

    Noise is a natural and unavoidable feature of real physical systems. In practical quantum systems, interactions with the surrounding environment lead to decoherence, which weakens the quantum coherence and reduces the fidelity of quantum state transfer. However, it has been observed that not all noise is entirely detrimental. Certain types of non-Markovian noise, which possess memory effects, can partially reverse the loss of coherence and even help restore quantum features that were previously degraded. Environmental noise becomes particularly problematic when scaling quantum walk systems to a large number of steps, making its study essential for realistic implementations.
    
    The influence of noise on continuous-time quantum walks has been explored in earlier work \cite{benedetti2016non, rossi2017continuous}. For example, in \cite{rossi2017continuous}, the authors examined how noise affects the localization behavior of quantum states on lattice graphs. In the context of discrete-time quantum walks, several studies have investigated the role of different noise models on various graph structures, such as path and cycle graphs \cite{chandrashekar2010relationship, chandrashekar2007symmetries, banerjee2017non}. More recently, a general framework for modelling discrete-time quantum walks in noisy environments, applicable to arbitrary graph topologies, was proposed in \cite{rani2024non}. Gaining insight into both noiseless and noisy quantum walk dynamics is important for identifying physical systems in which quantum walks can be reliably realized and controlled. 
    
    Quantum state transfer, benchmarked here using fidelity and coherence, is one of the key applications of quantum walks, and its performance is strongly influenced by environmental noise. As a result, it is essential to study state transfer in discrete-time quantum walks under realistic noisy conditions. In this work, we focus on unital non-Markovian Random Telegraph Noise (RTN) and a modified non-Markovian Ornstein–Uhlenbeck Noise (OUN) \cite{kumar2018non, banerjee2017non}, as well as a non-unital non-Markovian Amplitude Damping Noise (ADN) \cite{srikanth2008squeezed}. Although these noise models were originally introduced for qubits, they have been generalised to higher-dimensional systems using Weyl operators \cite{rani2026quantum}. Among the various ways to incorporate noise into quantum walks, we adopt an approach in which the quantum noise acts on the walker’s state after a finite number of quantum walk steps, following the method described in \cite{kumar2018non}. It is observed that when the sender and receiver are maximally separated, quantum communication is better than many other scenarios.
    
    The rest of the article is organised as follows. Section 2 outlines the relevant preliminaries from graph theory and provides a formal definition of the butterfly graphs considered in this work. In Section 3, we discuss the discrete-time quantum walks on arbitrary graphs. Further, the notion of quantum state transfer and coherence is explicated. Section 4 describes the discrete-time quantum walk on the butterfly graphs generated by different graphs. Section 5 introduces the Kraus operator formalism of unital RTN, OUN noises, and non-unital NMAD noise in higher dimensions utilised in this article. Section 6 presents the numerical results, focusing on fidelity behaviour and transfer efficiency in the absence of noise as well as under both unital and non-unital noise. Section 7 concludes the study by highlighting potential applications and describing future research directions.

\section{Preliminaries on graph theory}
    
    A graph $G = (V(G), E(G))$ is a combinatorial object consisting of a vertex set $V(G)$ and an edge set $E(G) \subset V(G) \times V(G)$ \cite{west2001introduction, bondy1976graph, dutta2025discrete, rahaman2026community}. The order of a graph is determined by its number of vertices. We depict a vertex by a dot, which may be marked or unmarked. An edge may be directed or undirected. An undirected edge $(u, v)$ is a line joining two vertices $u$ and $v$ that are the endpoints of the edge. We say $u$ and $v$ are the adjacent vertices. A directed edge $\overrightarrow{(u, v)}$ is a directed arc from the vertex $u$ to the vertex $v$. We say that the directed edge $\overrightarrow{(u, v)}$ is an outgoing edge from the vertex $u$ and an incoming edge to the vertex $v$. A graph without any directed edges is called an undirected graph. All the edges of a directed graph are directed. The simple undirected graphs do not have more than one edge between two vertices. Also, there is no edge in the simple graph joining a vertex with itself. We do not consider a weight on the vertices and the edges. In a simple undirected graph, the degree of a vertex $v$ is denoted by $d(v)$, which is the number of vertices adjacent to $v$. 
    
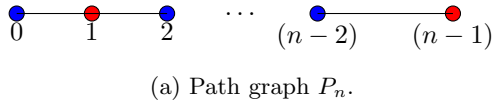
\begin{figure}
        \centering
        \begin{subfigure}[b]{.5\textwidth}
            \centering
            \begin{tikzpicture}
            \draw [fill = blue] (0, 0) circle [radius = 1mm];
            \node [below] at (0, 0) {$0$};
            \draw [fill = red] (1, 0) circle [radius = 1mm];
            \node [below] at (1, 0) {$1$};
            \draw [fill = blue] (2, 0) circle [radius = 1mm];
            \node [below] at (2, 0) {$2$};
            \draw (0, 0) -- (2, 0);
            \node at (3, 0) {$\dots$};
            \draw [fill = blue] (4, 0) circle [radius = 1mm];
            \node [below] at (4, 0) {$(n - 2)$};
            \draw [fill = red] (5.8, 0) circle [radius = 1mm];
            \node [below] at (5.8, 0) {$(n-1)$};
            \draw (4, 0) -- (5.8, 0);
            \end{tikzpicture}
            \caption{Path graph $P_n$.}
            \label{p_n}
        \end{subfigure}
        \caption{A path graph with $n$ vertices.}
        \label{simple_graph}
    \end{figure}

    A simple graph $G$ can also be represented by a directed graph $\overrightarrow{G}$ by adding two opposite orientations on every edge. Therefore, if $(u, v)$ be an unoriented edge in $G$, there are oriented edges $\overrightarrow{(u, v)}$ and $\overrightarrow{(v, u)}$ in the graph $\overrightarrow{G}$. Note that, if $G$ has $m$ undirected edges, then $\overrightarrow{G}$ has $2m$ directed edges. In this article, we use the path graphs $P_n$ of $n$ vertices, where we label the $n$ vertices with the integers $0, 1, \dots (n - 1)$. It has edges $(0, 1), (1, 2), \dots ((n-2), (n-1))$. \autoref{simple_graph} presents a path graph.

    A path in a graph between two vertices $s$ and $r$ is an alternating sequence of vertices and edges $ s = v_0, e_1, v_1, e_2, v_2, \\ \dots, e_l, v_l = r $. Here $l$ is the length of the path. The distance between $s$ and $r$ is the length of the shortest path between $s$ and $r$.
    
    A bipartite graph is a graph $G = (V(G), E(G))$ whose vertex set $V(G)$ can be partitioned into two subsets $V_1$ and $V_2$, such that every edge has an endpoint in $V_1$ and another in $V_2$. The subsets $V_1$ and $V_2$ are called the partite sets. In a graph $G$, a cycle is an alternating sequence of vertices and edges $\{v_0, e_1, v_1, e_2, v_2, \dots v_{n - 1}, e_n, v_n = v_0\}$, where no vertex is repeated except $v_0$. The cycle is called an odd or an even cycle, if $n$ is odd or even, respectively. It can be proved that a graph is a bipartite graph if and only if it has no odd cycle in it.
    
    Now, we introduce a family of butterfly graphs $\{B_0, B_1, \dots B_k\}$, which is constructed by the following procedure from a seed graph $P_n$.     
    \begin{procedure}\label{Butterfly_Procedure}
        Let $B_0 = B$, which is a graph with $n$ vertices, labeled by $0, 1, \dots (n - 1)$. For $k = 1, 2, \dots$ construct the graph $B_k$ as follows: 
        \begin{enumerate}
            \item 
                Take a copy of $B$ and label its vertices by the integers $(k + 1)n, (k + 1)n + 1, (k + 1)n + 2, \dots (k + 2)n - 1$.
            \item 
                Add the edges $(i, (k + 1)n + i)$ for $i = 0, 1, \dots (n - 1)$, where $i$ is a vertex of $B_0$ and $(k + 1)n + i$ is the corresponding vertex in the copy of $B$.
        \end{enumerate}
    \end{procedure}
    We call $B$ as the seed graph. In each step, we join a copy of $B$ with $B_0$ by joining corresponding vertices. If $B$ has no odd cycle, then any of the butterfly graphs $B_k$ also has no odd cycle. We consider the path graphs as our seeds in the examples. Hence, all the butterfly graphs considered in this article are bipartite graphs.
    
    A graph $H = (V(H), G(H))$ is said to be a subgraph of a graph $G = (V(G), E(G))$ if $V(H) \subset V(G)$ and $E(H) \subset E(G)$. The subgraph $H$ is said to be an induced subgraph if for $u, v \in V(H)$ and $(u, v) \in E(G)$ indicates $(u, v) \in E(H)$. We call the induced subgraphs generated by the copies of $B$ the wings in a butterfly graph $B_k$. The initial graph $B_0$ is the body of the butterfly $B_k$.
    
    \begin{example} 
        Now, we can construct a family of Butterfly graphs in a recursive way. Consider $B = P_2$ as a seed graph, which is the path graph with two vertices $0$ and $1$ and an edge $(0, 1)$. Therefore, the graph $B_0 = P_2$. To generate $B_1$ from $B_0$, we take another copy of $P_2$ and label the vertices with the integers $2, 3$ as well as add them with the corresponding vertices of $B_0$ by the edges $(0, 2)$ and $(1, 3)$. To construct $B_2$, we take a copy of $P_2$ and relabel its vertices by $4$ and $5$. Also, we add the edges $(0, 4)$ and $(1, 5)$. We continue the process similarly. The first few butterfly graphs $B_0, B_1, B_2$, and $B_{10}$ generated from $P_2$ are depicted in \autoref{butterfly_graphs_by_P2}. We also color the vertices in each $B_k$ in blue and red. The vertices of the same colors belong to the same partite set of a bipartite graph. 
    \end{example} 
    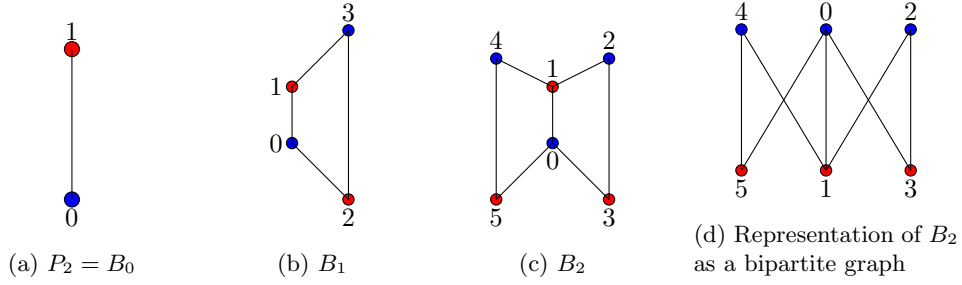
\begin{figure}
        \centering 
        \begin{subfigure}{.2\textwidth}
            \centering 
            \begin{tikzpicture} 
                \draw [fill = blue] (0, 0) circle [radius = 1mm];
                \node [below] at (0, 0) {${0}$};
                \draw [fill = red] (0, 2) circle [radius = 1mm];
                \node [above] at (0, 2) {${1}$};
                \draw (0, 0) -- (0, 2);
            \end{tikzpicture}
            \caption{$P_2 = B_0$}
        \end{subfigure}
        \begin{subfigure}{.15\textwidth}
            \centering
            \begin{tikzpicture}[scale = .75] 
                \draw [fill = blue] (0, 0) circle [radius = 1mm];
                \node [left] at (0, 0) {${0}$};
                \draw [fill = red] (0, 1) circle [radius = 1mm];
                \node [left] at (0, 1) {${1}$};
                \draw [fill = red] (1, -1) circle [radius = 1mm];
                \node [below] at (1, -1) {${2}$};
                \draw [fill = blue] (1, 2) circle [radius = 1mm];
                \node [above] at (1, 2) {${3}$};
                \draw (0, 0) -- (0, 1) -- (1, 2) -- (1, -1) -- (0, 0);
            \end{tikzpicture}
            \caption{$B_1$}
        \end{subfigure}
        \begin{subfigure}{0.2\textwidth}
            \centering
            \begin{tikzpicture}[scale = .75]
            \draw [fill = red] (0, 1) circle [radius = 1mm];
            \node [above] at (0, 1) {${1}$};
            \draw [fill = blue] (0, 0) circle [radius = 1mm];
            \node [below] at (0, 0) {${0}$};
            \draw [fill = blue] (1, 1.5) circle [radius = 1mm];
            \node [above] at (1, 1.5) {${2}$};
            \draw [fill = red] (1, -1) circle [radius = 1mm];
            \node [below] at (1, -1) {${3}$};
            \draw [fill = blue] (-1, 1.5) circle [radius = 1mm];
            \node [above] at (-1, 1.5) {${4}$};
            \draw [fill = red] (-1, -1) circle [radius = 1mm];
            \node [below] at (-1, -1) {${5}$};
            \draw (0, 1) -- (0, 0);
            \draw (0, 1) -- (1, 1.5);
            \draw (0, 1) -- (-1, 1.5);
            \draw (0, 0) -- (1, -1);
            \draw (0, 0) -- (-1, -1);
            \draw (1, 1.5) -- (1, -1);
            \draw (-1, 1.5) -- (-1, -1);
            \end{tikzpicture}
            \caption{$B_2$}
        \end{subfigure}
        \begin{subfigure}{0.2\textwidth}
            \centering
            \begin{tikzpicture}[scale = .75]
            \draw [fill = blue] (0, 1.5) circle [radius = 1mm];
            \node [above] at (0, 1.5) {${0}$};
            \draw [fill = red] (0, -1) circle [radius = 1mm];
            \node [below] at (0, -1) {${1}$};
            \draw [fill = blue] (1.5, 1.5) circle [radius = 1mm];
            \node [above] at (1.5, 1.5) {${2}$};
            \draw [fill = red] (1.5, -1) circle [radius = 1mm];
            \node [below] at (1.5, -1) {${3}$};
            \draw [fill = blue] (-1.5, 1.5) circle [radius = 1mm];
            \node [above] at (-1.5, 1.5) {${4}$};
            \draw [fill = red] (-1.5, -1) circle [radius = 1mm];
            \node [below] at (-1.5, -1) {${5}$};
            \draw (0, 1.5) -- (0, -1);
            \draw (0, 1.5) -- (1.5, -1);
            \draw (0, 1.5) -- (-1.5, -1);
            \draw (0, -1) -- (1.5, 1.5);
            \draw (0, -1) -- (-1.5, 1.5);
            \draw (1.5, 1.5) -- (1.5, -1);
            \draw (-1.5, 1.5) -- (-1.5, -1);
            \end{tikzpicture}    \caption{Representation of $B_2$ as a bipartite graph}     \label{Butterfly_to_bipartite}
        \end{subfigure}	
        \caption{Different butterfly graphs generated by the seed graph $P_2$. All these graphs are bipartite graphs. Vertices in two different partite sets are marked in two different colours.}
        \label{butterfly_graphs_by_P2}
    \end{figure}

\section{Discrete-time quantum walks on graphs}
    
    In quantum information theory, a pure quantum state is represented by a normalized vector in a Hilbert space, which we denote as a ket vector $\ket{\bullet}$. More generally, any quantum state can be described by a density matrix, which is a positive semidefinite, Hermitian matrix with unit trace. For a pure state $\ket{\psi}$, the corresponding density matrix is given by $\rho = \ket{\psi}\bra{\psi}$.
    
    In the quantum walks, \cite{banerjee2008symmetry, lovett2010universal}, a quantum walker moves across the vertices and edges of a graph following the topology of the graph. The evolution of the walker is governed by unitary operations that follow the rules of quantum mechanics. The state of the walker is described by a state vector in a Hilbert space, and its evolution is determined by a unitary operator, which is associated with the graph. In this article, we emphasize the discrete-time coined quantum walk, where the unitary evolution operator is the product of a coin operator and a shift operator. 
    
    In this work, we consider Grover's diffusion operators as the coin operators. The idea of this operator arises from Grover's search algorithm \cite{grover1996fast}. Let $G$ be a simple graph with $n$ vertices $0, 1, 2, \dots (n - 1)$. Let $d(i)$ be the degree of the vertex $i$ in the graph $G$. Hence, the vertex $i$ is adjacent to the vertices $j_1, j_2, \dots j_{d(i)}$. Corresponding to the vertices $j_1, j_2, \dots j_{d(i)}$, we assign the state vectors $\ket{j_1}, \ket{j_2}, \dots \ket{j_{d(i)}} \in \mathcal{H}^{(d(i))}$. Therefore, for every vertex $i$ we can define a $d(i)$-dimensional quantum state 
    \begin{equation}
        \ket{\phi(i)} =  \frac{1}{\sqrt{d(i)}} \sum_{(i,j) \in E(G)} \ket{j}.
    \end{equation}
    Now, the coin operator corresponding to the vertex $i$ is defined by 
    \begin{equation}\label{Grover_coin}
        C_i = 2 \ket{\phi(i)} \bra{\phi(i)} - I.
    \end{equation}
    In this article, we consider two marked vertices $s$ and $r$ in the graph, which are assigned for the sender and the receiver of the quantum information, respectively. We multiply $-1$ with the coin operators corresponding to the marked vertices. The coin operator for the movement of the walker on the graph is determined by
    \begin{equation}\label{coin}
        C = C_0 \bigoplus C_1 \bigoplus \dots \bigoplus (- C_r) \bigoplus \dots \bigoplus (- C_s) \bigoplus \dots \bigoplus C_{n - 1},
    \end{equation}
    where $\bigoplus$ denotes the direct sum of matrices. As $C_i$ is a matrix of order $d(i)$, order of the matrix $C$ is $\sum_{i = 0}^{n - 1} d(i) = 2m$, where $m$ is the number of edges in $G$. 
    
    Now, to define the shift operator, we convert the simple graph $G$ to a directed graph $\overrightarrow{G}$ by assigning two opposite orientations on every undirected edge. Recall that the number of edges in $\overrightarrow{G}$ is $2m$. Corresponding to a directed edge $\overrightarrow{(i, j)}$ we assign a state vector $\ket{\overrightarrow{(i, j)}} \in \mathcal{H}^{2m}$. The shift operator in a DTQW governs the movement of the walker on the directed edges of the graph. The shift operator $S$ is defined by
    \begin{equation}\label{shift}
    S\ket{\overrightarrow{(i,j)}} = \ket{\overrightarrow{(j,i)}}.
    \end{equation}
    Therefore, the operator $S$ swaps the direction of the walker from the edge  $\overrightarrow{(i,j)}$ to $\overrightarrow{(j, i)}$. Note that the order of $S$ is $2m$.
    
    After deriving the coin and the shift operator of the quantum walk, we generate a unitary operator $U = SC$. As both $S$ and $C$ are matrices of order $2m$, the order of the matrix $U$ is also $2m$. 
    
    We have assumed that a sender located at the vertex $s$ sends quantum information via the structure of graphs. Let the outgoing edges from the vertex $s$ be $\overrightarrow{(s, t_1)}, \overrightarrow{(s, t_2)}, \dots, \overrightarrow{(s, t_{d(s)})}$. We can construct a quantum state 
    \begin{equation}\label{sender_state}
        \ket{\psi(s)} = \frac{1}{\sqrt{d(s)}} \left[\ket{\overrightarrow{(s, t_1)}} + \ket{\overrightarrow{(s, t_2)}} + \dots + \ket{\overrightarrow{(s, t_{d(s)})}} \right].
    \end{equation}
    Similarly, let the incoming edges to the vertex $r$ be $\overrightarrow{(q_1, r)}, \overrightarrow{(q_2, r)}, \dots \overrightarrow{(q_{d(r)}, r)}$ then we can construct a quantum state
    \begin{equation}\label{receiver_state}
        \ket{\psi(r)} = \frac{1}{\sqrt{d(r)}} \left[ \ket{\overrightarrow{(q_1, r)}} + \ket{\overrightarrow{(q_2, r)}} + \dots + \ket{\overrightarrow{(q_{d(r)}, r)}}\right].
    \end{equation}
    We consider that the state of the sender $\ket{\psi(s)}$ is the initial state of the quantum walker at $t = 0$. At time $t$, the state of the walker is 
    \begin{equation}
        \ket{\psi_t} = U^t \ket{\psi(s)}.
    \end{equation} 
    We say that there is a perfect state transfer from the vertex $s$ to the vertex $r$ at time $t = k$ if $\ket{\psi_k} = \ket{\psi(r)}$. Perfect state transfer is a rare phenomenon on simple graphs. 
    
    To investigate state transfer in quantum walks, fidelity \cite{liang2019quantum} is employed as a benchmark to compare the time-evolved state with the target receiver state. To find the similarity between $\ket{\psi_t}$ and $\ket{\psi(r)}$, we calculate fidelity between them, which is defined by
    \begin{equation}\label{Fidelity_vectors}
        F(\ket{\psi_t}, \ket{\psi(r)}) = |\bra{\psi_t} \ket{\psi(r)}|^2.
    \end{equation}
    A quantum state can also be represented by a density matrix. The fidelity between two quantum states described by density matrices $\rho$ and $\sigma$ is defined as
    \begin{equation}\label{fidelity_density}
        F(\rho, \sigma) = \left( \text{Tr} \left[ \sqrt{ \sqrt{\rho} \sigma \sqrt{\rho} } \right] \right)^2.
    \end{equation}
    Fidelity takes values from $0$ to $1$, where $0$ indicates complete dissimilarity and $1$ represents perfect overlap. This measure provides a precise way to evaluate how effectively a quantum state has been transferred or evolved into the desired state. In our work, we examine the fidelity of state transfer between two locations on a graph at different time steps, and determine which locations are the best among all possible locations. To determine it quantitatively, we take the time average of fidelity values. If at time $t$ fidelity is $F_t$ for $t = 1, 2, \dots$, then we define the average fidelity by    \begin{equation}\label{Average_fidelity}
        \mathcal{F}_T = \frac{1}{T} \sum_{t = 1}^T F_t
    \end{equation}. 
    
    Decoherence dynamics is a central aspect in the study of open quantum systems. Quantum decoherence refers to the degradation of superposition due to interactions with the environment. In this work, we quantify coherence \cite{chen2016quantifying, baumgratz2014quantifying} using the $l_1$
	norm, which captures the contribution of off-diagonal elements of the density matrix, is given as
    \begin{equation}\label{coherence}
	C_{l_1}(\rho) = \sum_{i\neq j}|\rho_{ij}|.
	\end{equation}

\section{Discrete-time quantum walk on the butterfly graphs}
\label{Discrete-time_quantum_walk_on_the_butterfly_graphs}
    
    We consider that there is a sender and a receiver placed at two distinct vertices of a butterfly graph. We mark these vertices with $s$ and $r$, respectively. It classifies the vertex set $V(B_k)$ into the marked and unmarked vertices. We consider different cases based on the locations of the marked vertices.

    \subsection{State transfer on the butterfly graphs generated by $P_2$}

        As we initiate our process with $P_2$, it is the first element of our family of butterfly graphs. Therefore, $B_0 = P_2$. We consider $P_2$ as a seed graph in our network generating process. We construct some members of this family, which are depicted in \autoref{butterfly_graphs_by_P2}.

        \subsubsection{The seed graph $P_2$}
\begin{figure}
            \centering 
            \begin{subfigure}{.2\textwidth}
                \begin{subfigure}{.15\textwidth}
                    \centering
                    \begin{tikzpicture} 
                        \draw [fill = blue] (0, 0) circle [radius = 1mm];
                        \node [below] at (0, 0) {${0} (s)$};
                        \draw [fill = red] (0, 2) circle [radius = 1mm];
                        \node [above] at (0, 2) {${1} (r)$};
                        \draw (0, 0) -- (0, 2);
                    \end{tikzpicture}
                    \caption{$P_2$}
                    \label{B0_by_P2}
                \end{subfigure}
            \end{subfigure}
            \begin{subfigure}{0.5\textwidth}
                \includegraphics[width=\textwidth]{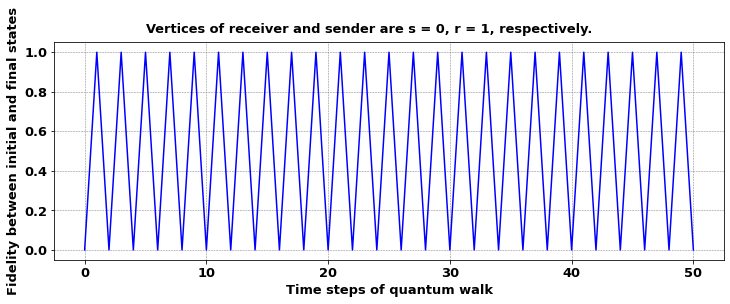}
                \caption{Fidelity at different steps of quantum walk.}
                \label{Fidelity_P_2}
            \end{subfigure}
            \caption{State transfer on $P_2$.}
        \end{figure}
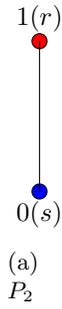
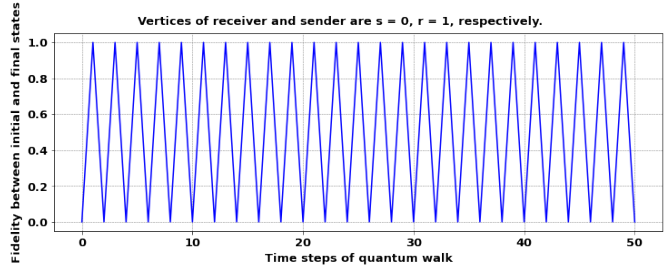

        The graph $P_2$ consists of two vertices $0$ and $1$ and an edge, depicted in \autoref{B0_by_P2}. We can place the sender and receiver interchangeably at $1$ and $0$, respectively. Clearly $\ket{\phi(s)} = \ket{1}$ and $\ket{\phi(r)} = \ket{0}$. The shift, coin, and evaluation operators leading the quantum walk are represented by
        \begin{equation}
            S_{P_2} = \begin{bmatrix}
                    0 & 1\\
                    1 & 0
                \end{bmatrix}, 
            C_{P_2} = \begin{bmatrix}
                    -1 & 0\\
                    0 & -1
                \end{bmatrix}, ~\text{and}~
                U_{P_2} = S_{P_2}C_{P_2} = \begin{bmatrix}
                    0 & -1\\
                    -1 & 0
                \end{bmatrix}, ~\text{respectively}.
        \end{equation}
        The fidelity values in different steps of the quantum walk are depicted in \autoref{Fidelity_P_2}. It is observed that the fidelity to transfer the state from the sender to receiver at $t = 1, 3, \dots$ is $1$. It matches the observations in \cite{rani2026quantum}.

        \subsubsection{The butterfly $B_1$ generated by $P_2$}
        In the next step, we increase our network using the butterfly product by adding a wing. The graph $B_1$ has four vertices, which are depicted in \autoref{B1_by_P2}. The shift operator depends on the edges in the graph. For $B_1$, the shift operator is given by 
        \begin{equation}
            S_{B_1} = \begin{bmatrix}
                0 & 0 & 1 & 0 & 0 & 0 & 0 & 0 \\
                0 & 0 & 0 & 0 & 0 & 1 & 0 & 0 \\
                1 & 0 & 0 & 0 & 0 & 0 & 0 & 0 \\
                0 & 0 & 0 & 0 & 0 & 0 & 0 & 1 \\
                0 & 0 & 0 & 0 & 0 & 0 & 1 & 0 \\
                0 & 1 & 0 & 0 & 0 & 0 & 0 & 0 \\
                0 & 0 & 0 & 0 & 1 & 0 & 0 & 0 \\
                0 & 0 & 0 & 1 & 0 & 0 & 0 & 0
            \end{bmatrix}.
        \end{equation}
        When the degree of a vertex is two, the Grover operator becomes the Pauli $X$ operator, which is $X = \begin{bmatrix} 0 & 1 \\ 1 & 0 \end{bmatrix}.$
        Now, the sender and receiver may be kept in different vertices of the network. Due to the locations of the sender and receiver, we have the following cases:
        \begin{enumerate}
            \item 
                \textbf{The sender and receivers are placed on the body of the butterfly.}\\ 
                Let the sender be placed at vertex $s = 0$, and the receiver be placed at vertex $r = 1$. From equation \eqref{sender_state}, and \eqref{receiver_state} we have
                \begin{equation}
                        \ket{\psi(s)} = \frac{1}{\sqrt{2}} \left[\ket{\overrightarrow{(0, 1)}} + \ket{\overrightarrow{(0, 2)}}\right]; \text{and}~ \ket{\psi(r)} = \frac{1}{\sqrt{2}} \left[\ket{\overrightarrow{(1, 0)}} + \ket{\overrightarrow{(1, 3)}}\right].
                \end{equation}
                The evolution operator $U_{B1} = S_{B_1}C_{B_1}$ where $C_{B_1}$ is a block diagonal matrix represented by $C_{B_1} = \diag\{-X, -X, X, X\}$, in this case. We observe that the fidelity becomes $0.25$ at $t = 1, 3, 5, \dots$. The fidelity plot is shown in \autoref{Fidelity_P2_B1_01}. When the sender is at the vertex $2$, and the receiver is at the vertex $3$, we have a similar observation. 
            
            \item 
                \textbf{One of the sender and receiver is placed on the body and the other is placed on the wing of the butterfly, and they are on the same partite set.} \\
                Let the sender and receiver be placed at vertices $s = 1$ and $r = 2$, respectively. Clearly,
                \begin{equation}
                    \ket{\psi(s)} = \frac{1}{\sqrt{2}} \left[\ket{\overrightarrow{(1, 0)}} + \ket{\overrightarrow{(1, 3)}}\right]; ~\text{and}~ \ket{\psi(r)} = \frac{1}{\sqrt{2}} \left[\ket{\overrightarrow{(2, 0)}} + \ket{\overrightarrow{(2, 3)}}\right].
                \end{equation}
                The coin operator is given by $C_{B_1} = \diag\{X, -X, -X, X\}$.
                We find that the fidelity of state transfer is $1$ at $t = 2, 4, 6, \dots$ as depicted in \autoref{Fidelity_P2_B1_12}. The same result is observed for the other case when the sender is at $s = 0$ and $r = 3$.
            \item 
                \textbf{One of the sender and receiver is placed on the body and the other is placed on the wing of the butterfly, as well as they are in the different partite set.}\\
                Let the sender $s = 0$ and $r = 2$. Clearly,
                \begin{equation}
                    \ket{\psi(s)} = \frac{1}{\sqrt{2}} \left[\ket{\overrightarrow{(0, 1)}} + \ket{\overrightarrow{(0, 2)}}\right]; ~\text{and}~ \ket{\psi(r)} = \frac{1}{\sqrt{2}} \left[\ket{\overrightarrow{(2, 0)}} + \ket{\overrightarrow{(2, 3)}}\right]
                \end{equation}
                The coin operator is given by $C_{B_1} = \diag\{-X, X, -X, X\}$. Our numerical observations suggest that the maximum fidelity is $0.25$ for $t = 1, 3, 5 \dots$. As it is similar to Case 1, we omit the corresponding graph.
        \end{enumerate}  
        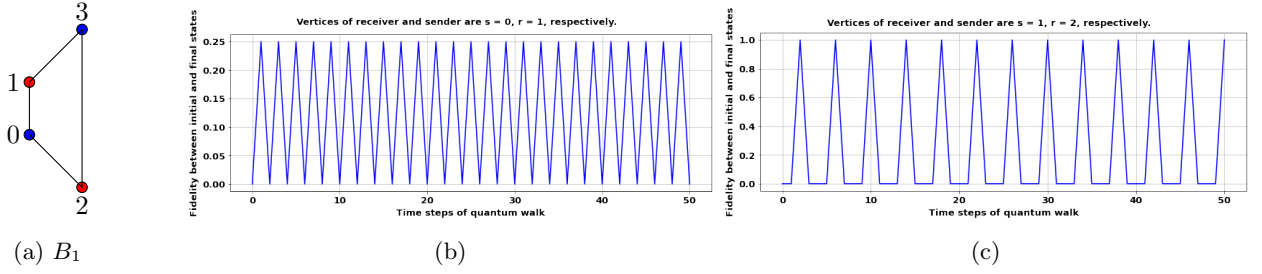
\begin{figure}
            \begin{subfigure}{.2\textwidth}
                \centering
                \begin{tikzpicture}[scale = .7] 
                    \draw [fill = blue] (0, 0) circle [radius = 1mm];
                    \node [left] at (0, 0) {${0}$};
                    \draw [fill = red] (0, 1) circle [radius = 1mm];
                    \node [left] at (0, 1) {${1}$};
                    
                    \draw [fill = red] (1, -1) circle [radius = 1mm];
                    \node [below] at (1, -1) {${2}$};
                    \draw [fill = blue] (1, 2) circle [radius = 1mm];
                    \node [above] at (1, 2) {${3}$};
                    \draw (0, 0) -- (0, 1) -- (1, 2) -- (1, -1) -- (0, 0);
                \end{tikzpicture}
                \caption{$B_1$}
                \label{B1_by_P2}
            \end{subfigure}
            \begin{subfigure}{0.4\textwidth}
                \includegraphics[width=\textwidth]{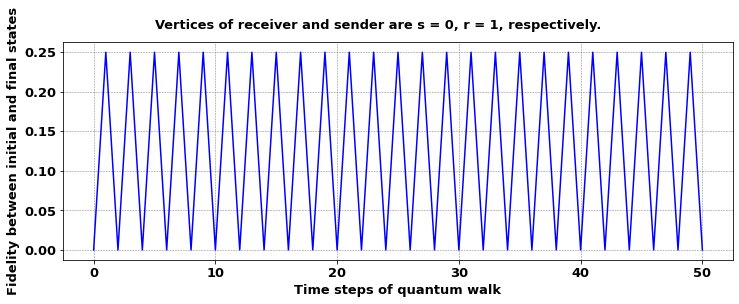}
                \caption{}
                \label{Fidelity_P2_B1_01}
            \end{subfigure}
            \begin{subfigure}{0.4\textwidth}
                \includegraphics[width=\textwidth]{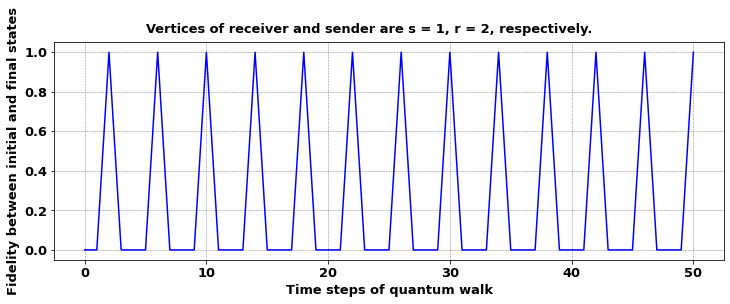}
                \caption{}
                \label{Fidelity_P2_B1_12}
            \end{subfigure}
            \caption{State transfer on the butterfly graph $B_1$.}
        \end{figure}

        From the above three cases, we observe that in $B_1$, the maximum fidelity occurs when one of the sender and receiver is placed on the body, and the other is placed on the wing of the butterfly. Also, they must be on the same partite set. In the other cases, the fidelity is limited to $0.25$. The average fidelity values for a few locations of sender and receiver are shown in \autoref{table_B_1_P_2}. Note that the maximum possible distance between two vertices in $B_1$ is $2$, which is the distance between the vertices $1$ and $2$.

        \begin{table}[hbt!]
        \centering
        \small
        \begin{tabular}{|c|c|c|}
        \hline
        \textbf{Location of Sender} & \textbf{Receiver Node} & \textbf{Average Fidelity} \\
        \hline
         0 & 1 & 0.125 \\
        \hline
         \textbf{1} & \textbf{2} & \textbf{0.25} \\
        \hline
         0 & 2 & 0.125 \\
        \hline
        \end{tabular}
        \caption{Average fidelity for different sender and receiver locations up to $200$ time-steps on the butterfly graph $B_{1}$ generated by path graph $P_{2}$.}
        \label{table_B_1_P_2}
        \end{table}
        
        \subsubsection{The butterfly graph $B_2$ generated by $P_2$}
\begin{figure}
         \centering 
        \begin{subfigure}{.32\textwidth}
            \centering
            \begin{tikzpicture}[scale = .7] 
            \draw [fill = blue] (0, 0) circle [radius = 1mm];
            \node [left] at (0, 0) {${0}$};
            \draw [fill = red] (0, 1) circle [radius = 1mm];
            \node [left] at (0, 1) {${1}$};
            
            \draw [fill = red] (1, -1) circle [radius = 1mm];
            \node [below] at (1, -1) {${2}$};
            \draw [fill = blue] (1, 2) circle [radius = 1mm];
            \node [above] at (1, 2) {${3}$};
            
            \draw [fill = red] (-1, -1) circle [radius = 1mm];
            \node [below] at (-1, -1) {${4}$};
            \draw [fill = blue] (-1, 2) circle [radius = 1mm];
            \node [above] at (-1, 2) {${5}$};
            
            \draw (0, 0) -- (0, 1) -- (1, 2) -- (1, -1) -- (0, 0) -- (-1, -1) -- (-1, 2) -- (0, 1);
            \end{tikzpicture}
            \caption{$B_2$}
            \label{B2_by_P2}
        \end{subfigure}
        \begin{subfigure}{0.32\textwidth}
            \includegraphics[width=\textwidth]{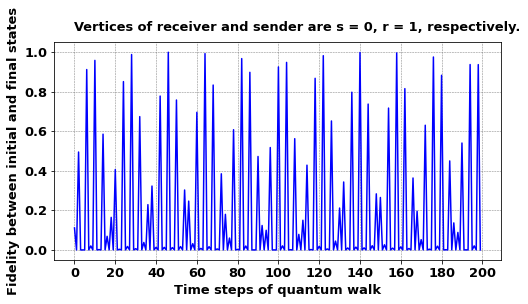}
            \caption{}
            \label{P2_B2_01_fidelity}
        \end{subfigure}
       \begin{subfigure}{0.32\textwidth}
        \includegraphics[width=\textwidth]{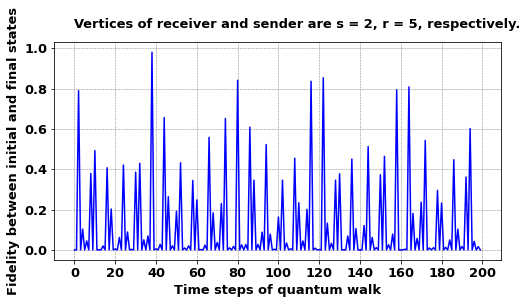}
        \caption{}
        \label{P2_B2_5_fidelity}
       \end{subfigure}
       \\
       \begin{subfigure}{0.32\textwidth}
        \includegraphics[width=\textwidth]{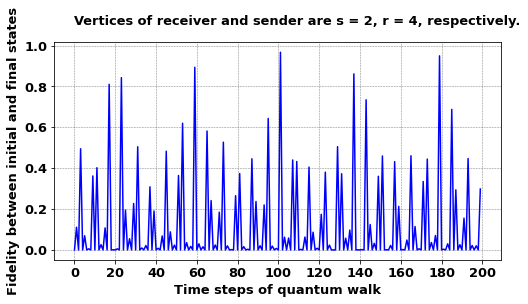}
        \caption{}
        \label{P2_B2_24_fidelity}
       \end{subfigure}
       \begin{subfigure}{0.32\textwidth}
        \includegraphics[width=\textwidth]{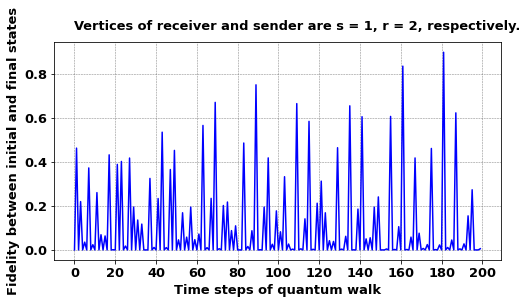}
        \caption{}
        \label{P2_B2_12_fidelity}
       \end{subfigure}
       \begin{subfigure}{0.32\textwidth}
        \includegraphics[width=\textwidth]{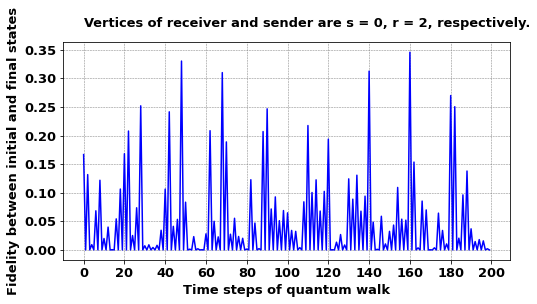}
        \caption{}
        \label{P2_B2_02_fidelity}
       \end{subfigure}
        \caption{State transfer on the butterfly graph $B_2$.}
       \end{figure}
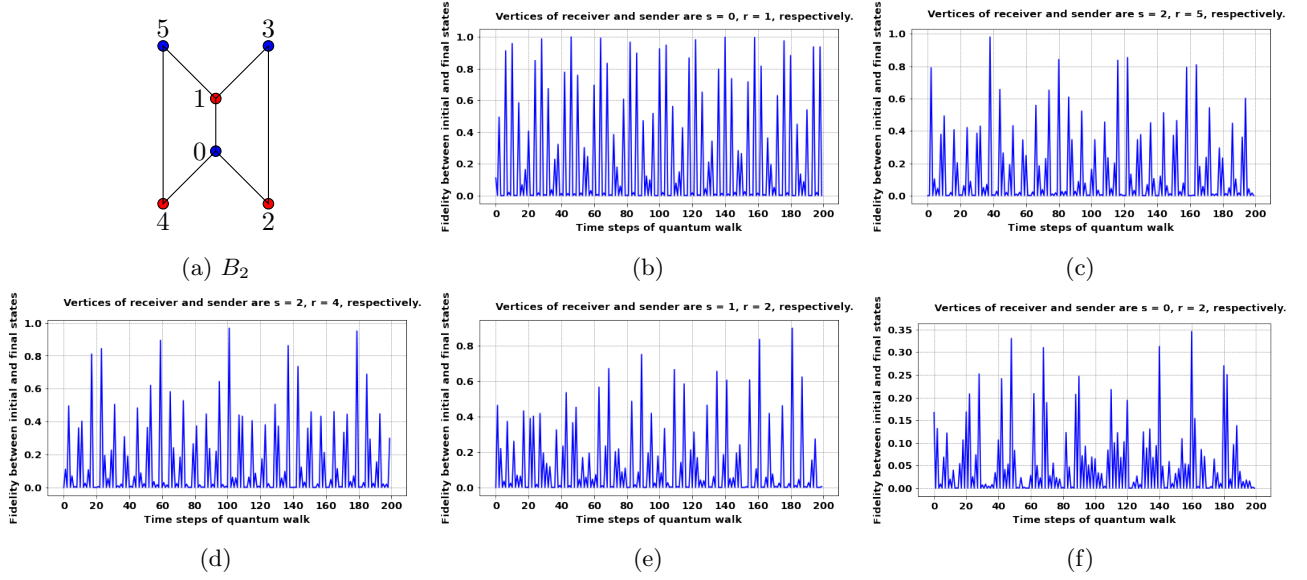

         Now, we increase our network by adding a wing to $B_1$ to form $B_2$. The graph $B_2$ has six vertices, which are depicted in \autoref{B2_by_P2}. The shift operator for $B_2$ is given by
         \begin{equation}
            S_{B_2} = \begin{bmatrix}
                0 & 0 & 0 & 1 & 0 & 0 & 0 & 0 & 0 & 0 & 0 & 0 & 0 & 0 \\
                0 & 0 & 0 & 0 & 0 & 0 & 0 & 1 & 0 & 0 & 0 & 0 & 0 & 0 \\
                0 & 0 & 0 & 0 & 0 & 0 & 0 & 0 & 0 & 0 & 0 & 1 & 0 & 0 \\
                1 & 0 & 0 & 0 & 0 & 0 & 0 & 0 & 0 & 0 & 0 & 0 & 0 & 0 \\
                0 & 0 & 0 & 0 & 0 & 0 & 0 & 0 & 0 & 1 & 0 & 0 & 0 & 0 \\
                0 & 0 & 0 & 0 & 0 & 0 & 0 & 0 & 0 & 0 & 0 & 0 & 0 & 1 \\
                0 & 0 & 0 & 0 & 0 & 0 & 0 & 0 & 1 & 0 & 0 & 0 & 0 & 0 \\
                0 & 1 & 0 & 0 & 0 & 0 & 0 & 0 & 0 & 0 & 0 & 0 & 0 & 0 \\
                0 & 0 & 0 & 0 & 0 & 0 & 1 & 0 & 0 & 0 & 0 & 0 & 0 & 0 \\
                0 & 0 & 0 & 0 & 1 & 0 & 0 & 0 & 0 & 0 & 0 & 0 & 0 & 0 \\
                0 & 0 & 0 & 0 & 0 & 0 & 0 & 0 & 0 & 0 & 0 & 0 & 1 & 0 \\
                0 & 0 & 1 & 0 & 0 & 0 & 0 & 0 & 0 & 0 & 0 & 0 & 0 & 0 \\
                0 & 0 & 0 & 0 & 0 & 0 & 0 & 0 & 0 & 0 & 1 & 0 & 0 & 0 \\
                0 & 0 & 0 & 0 & 0 & 1 & 0 & 0 & 0 & 0 & 0 & 0 & 0 & 0
            \end{bmatrix}.
          \end{equation}
        Placing the sender and receivers at different vertices of the graph, we have the following cases:
        \begin{enumerate}
            \item 
            \textbf{The sender and receivers are placed on the body of the butterfly.} \\
            Let the sender and the receiver be located at $s = 0$ and $r = 1$. From equation \eqref{sender_state} and \eqref{receiver_state} we have
            \begin{equation}
            \ket{\psi(s)} = \frac{1}{\sqrt{3}} \left[\ket{\overrightarrow{(0, 1)}} + \ket{\overrightarrow{(0, 2)}} + \ket{\overrightarrow{(0, 4)}}\right]; \text{and}~ \ket{\psi(r)} = \frac{1}{\sqrt{3}} \left[\ket{\overrightarrow{(1, 0)}} + \ket{\overrightarrow{(1, 3)}} +\ket{\overrightarrow{(1, 5)}}\right].
            \end{equation}
            The following equation shows the coin operator for this case:
            \begin{equation}
                C = -C_0 \bigoplus - C_1 \bigoplus C_2 \bigoplus C_3 \bigoplus C_4 \bigoplus C_5 ,
            \end{equation}
            where 
            \begin{equation} \label{Coin_3}
            C_{0} = C_{1}= \begin{bmatrix}
                -0.33333333 & 0.66666667 & 0.66666667\\
                 0.66666667 & -0.33333333 & 0.66666667\\
                 0.66666667 & 0.66666667 & -0.33333333
            \end{bmatrix},
            \end{equation}
            and  \begin{equation} \label{Coin_2}
            C_{2} = C_{3}= C_{4}= C_{5} = \begin{bmatrix}
                0 & 1\\
                1 & 0
            \end{bmatrix},
            \end{equation}
            are determined by equation \eqref{Grover_coin}. As depicted in \autoref{P2_B2_01_fidelity}, the fidelity is more than $0.8$ at time steps $t = 7, 11, 25, 29, 43, 47, \dots$. 
            \item 
            \textbf{The sender and receivers are placed on the wings of the butterfly with the maximum distance, and they are in different partite sets.} 
            
            Let $s = 2$ and $r = 5$. Then,
            \begin{equation}
            \ket{\psi(s)} = \frac{1}{\sqrt{2}} \left[\ket{\overrightarrow{(2, 0)}} + \ket{\overrightarrow{(2, 3)}} \right]; \text{and}~ \ket{\psi(r)} = \frac{1}{\sqrt{2}} \left[\ket{\overrightarrow{(5, 1)}} + \ket{\overrightarrow{(5, 4)}} \right].
            \end{equation}
            For this case, the coin operator is given by
            \begin{equation}
                C = C_0 \bigoplus  C_1 \bigoplus -C_2 \bigoplus C_3 \bigoplus C_4 \bigoplus -C_5 .
            \end{equation}

            As depicted in \autoref{P2_B2_5_fidelity}, state transfer with fidelity more than $0.8$ is achieved at time steps $t = 7, 39, 81, \dots$. The identical results are obtained for the case when $s = 3$ and $r = 4$.
            
            \item 
            \textbf{The sender and receiver are placed on the wings of the butterfly with the maximum distance, and they are in the same partite set.}
            
            Next we assume, $s = 2$ and $r = 4$.
            \begin{equation}
            \ket{\psi(s)} = \frac{1}{\sqrt{2}} \left[\ket{\overrightarrow{(2, 0)}} + \ket{\overrightarrow{(2, 3)}} \right]; \text{and}~ \ket{\psi(r)} = \frac{1}{\sqrt{2}} \left[\ket{\overrightarrow{(4, 0)}} + \ket{\overrightarrow{(4, 5)}} \right].
            \end{equation}
            The coin operator for this case is given by
            \begin{equation}
                C = C_0 \bigoplus  C_1 \bigoplus -C_2 \bigoplus C_3 \bigoplus -C_4 \bigoplus C_5.
            \end{equation}
             \autoref{P2_B2_24_fidelity} shows that the fidelity of state transfer is more than $0.8$ at time steps $ t= 18, 24, 60, 102, \dots$.  
            \item 
            \textbf{One of the sender and receivers is placed on the body, and the other is placed on the wing of the butterfly, and they are on the same partite set.}
            
            Here we consider $s = 1$ and $r = 2$. Clearly,
            \begin{equation}
            \ket{\psi(s)} = \frac{1}{\sqrt{3}} \left[\ket{\overrightarrow{(1, 0)}} + \ket{\overrightarrow{(1, 3)}} + \ket{\overrightarrow{(1, 5)}}\right]; \text{and}~ \ket{\psi(r)} = \frac{1}{\sqrt{2}} \left[\ket{\overrightarrow{(2, 0)}} + \ket{\overrightarrow{(2, 3)}} \right].
            \end{equation} 
            The coin operator is given by
            \begin{equation}
                C = C_0 \bigoplus  -C_1 \bigoplus -C_2 \bigoplus C_3 \bigoplus C_4 \bigoplus C_5.
            \end{equation}
            The fidelity of state transfer is found to be more than $0.8$ at time steps $ t = 162, 182, \dots $ as illustrated in \autoref{P2_B2_12_fidelity}. 
            \item 
            \textbf{One of the sender and receivers is placed on the body, and the other is placed on the wing of the butterfly, and they are in a different partite set.}
            
            Assume that the sender and the receiver are at $s = 0$ and $r = 2$. Therefore, 
            \begin{equation}
            \ket{\psi(s)} = \frac{1}{\sqrt{3}} \left[\ket{\overrightarrow{(0, 1)}} + \ket{\overrightarrow{(0, 2)}} + \ket{\overrightarrow{(0, 4)}}\right]; \text{and}~ \ket{\psi(r)} = \frac{1}{\sqrt{2}} \left[\ket{\overrightarrow{(2, 0)}} + \ket{\overrightarrow{(2, 3)}} \right].
            \end{equation}
            The coin operator is described by using equation \eqref{coin},
            \begin{equation}
                C = -C_0 \bigoplus  C_1 \bigoplus -C_2 \bigoplus C_3 \bigoplus C_4 \bigoplus C_5.
            \end{equation}
            We observe that the fidelity of state transfer is $0.34$ at time step $t =161 $, which is the maximum fidelity up to $200$ time steps, as shown in  \autoref{P2_B2_02_fidelity}.
        \end{enumerate}

       \subsubsection{The butterfly graph $B_3$ generated by $P_2$}
       
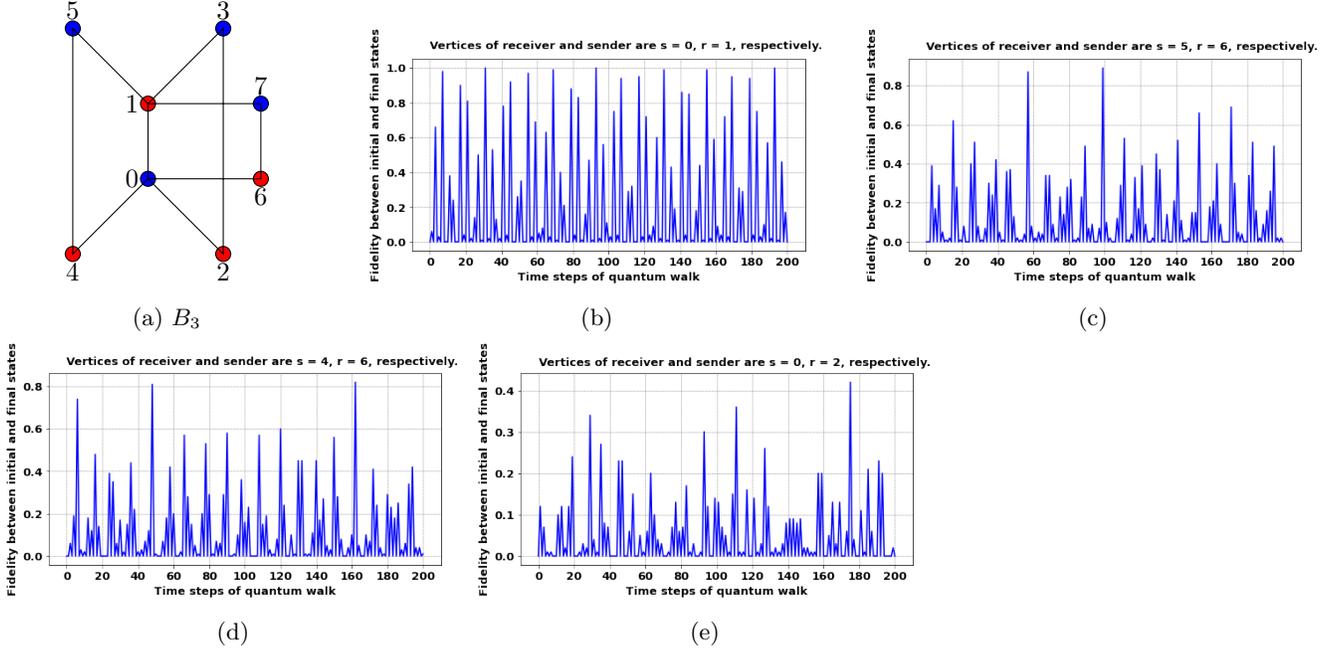
\begin{figure}
            \begin{subfigure}{.25\textwidth}
                \centering
                \begin{tikzpicture} 
                \draw [fill = blue] (0, 0) circle [radius = 1mm];
                \node [left] at (0, 0) {${0}$};
                \draw [fill = red] (0, 1) circle [radius = 1mm];
                \node [left] at (0, 1) {${1}$};
                
                \draw [fill = red] (1, -1) circle [radius = 1mm];
                \node [below] at (1, -1) {${2}$};
                \draw [fill = blue] (1, 2) circle [radius = 1mm];
                \node [above] at (1, 2) {${3}$};
                
                \draw [fill = red] (-1, -1) circle [radius = 1mm];
                \node [below] at (-1, -1) {${4}$};
                \draw [fill = blue] (-1, 2) circle [radius = 1mm];
                \node [above] at (-1, 2) {${5}$};
                
                \draw [fill = red] (1.5, 0) circle [radius = 1mm];
                \node [below] at (1.5, 0) {${6}$};
                \draw [fill = blue] (1.5, 1) circle [radius = 1mm];
                \node [above] at (1.5, 1) {${7}$};
                
                \draw (0, 0) -- (0, 1) -- (1, 2) -- (1, -1) -- (0, 0) -- (-1, -1) -- (-1, 2) -- (0, 1) -- (1.5, 1) -- (1.5, 0) -- (0, 0);
                \end{tikzpicture}
                \caption{$B_{3}$}
                \label{B3_by_P2}
            \end{subfigure}
            \begin{subfigure}{0.35\textwidth}
            \includegraphics[width=\textwidth]{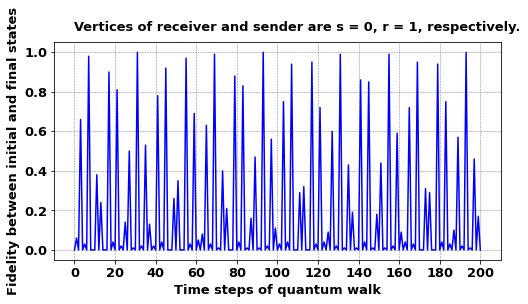}
                \caption{}
                \label{P2_B3_01_fidelity}
            \end{subfigure}
            \begin{subfigure}{0.35\textwidth}
            \includegraphics[width=\textwidth]{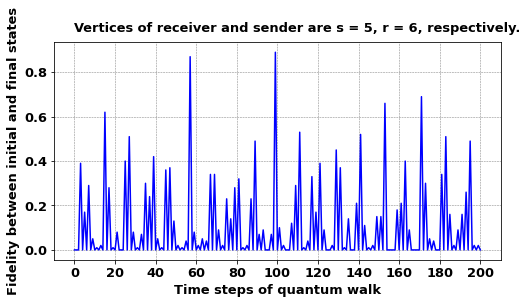}
                \caption{}
                \label{P2_B3_56_fidelity}
            \end{subfigure}
            \begin{subfigure}{0.35\textwidth}
            \includegraphics[width=\textwidth]{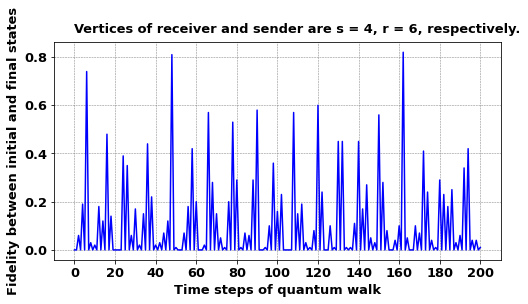}
                \caption{}
                \label{P2_B3_46_fidelity}
            \end{subfigure}
            \begin{subfigure}{0.35\textwidth}
            \includegraphics[width=\textwidth]{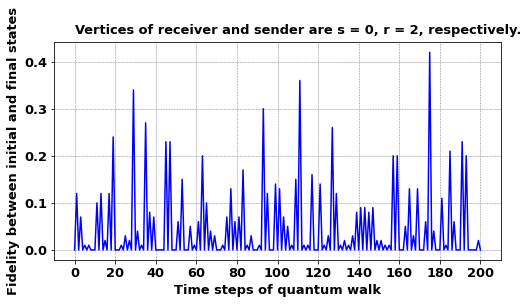}
                \caption{}
                \label{P2_B3_02_fidelity}
            \end{subfigure}
            \caption{State transfer on the butterfly graph $B_3$ generated by the path graph $P_2$.}
        \end{figure}

        Next, we construct $B_3$ by adding one more wing to $B_2$, which is depicted in \autoref{B3_by_P2}. We observe a number of cases where the sender and receiver are in different parts of the network.
        \begin{enumerate}
            \item 
            \textbf{The sender and receivers are placed on the body of the butterfly}. 
            
            Let the sender and the receiver are at $s = 0$ and $r = 1$ respectively. From equation \eqref{sender_state} and \eqref{receiver_state}, we have
            \begin{equation}
            \begin{split}
            \ket{\psi(s)} = & \frac{1}{\sqrt{4}} \left[\ket{\overrightarrow{(0, 1)}} + \ket{\overrightarrow{(0, 2)}} + \ket{\overrightarrow{(0, 4)}} + \ket{\overrightarrow{(0, 6)}}\right], \\
            \ket{\psi(r)} = & \frac{1}{\sqrt{4}} \left[\ket{\overrightarrow{(1, 0)}} + \ket{\overrightarrow{(1, 3)}} +\ket{\overrightarrow{(1, 5)}} + \ket{\overrightarrow{(1, 7)}}\right].
            \end{split}
            \end{equation}
            and from equation \eqref{coin} the coin operator is given by
            \begin{equation}
                C = -C_0 \bigoplus  -C_1 \bigoplus C_2 \bigoplus C_3 \bigoplus C_4 \bigoplus C_5 \bigoplus C_6 \bigoplus C_7,
            \end{equation}
            where the coin operator for the vertices with degree $4$ are defined as
            \begin{equation}\label{Coin_4}
                C_{0} = C_{1} =\begin{bmatrix}
                    -0.5 & 0.5 & 0.5 & 0.5\\
                    0.5 & -0.5 & 0.5 & 0.5\\
                    0.5 & 0.5 & -0.5 & 0.5\\
                    0.5 & 0.5 & 0.5 & -0.5
                \end{bmatrix}
            \end{equation}
            and the coin operator for all the vertices with a degree two is as in equation \eqref{Coin_2}. The fidelity of state transfer is found to be equal to $1$ at time steps $ t = 31, 69, 93, \dots $ as illustrated in  \autoref{P2_B3_01_fidelity}.        
            \item 
            \textbf{The sender and receivers are placed on the wings of the butterfly with the maximum distance, and they are in different partite set.} 
            
            Now we assume that $s = 5$ and $r = 6$. Similarly,
            \begin{equation}
            \ket{\psi(s)} = \frac{1}{\sqrt{2}} \left[\ket{\overrightarrow{(5, 1)}} + \ket{\overrightarrow{(5, 4)}} \right]; \text{and}~ \ket{\psi(r)} = \frac{1}{\sqrt{2}} \left[\ket{\overrightarrow{(6, 0)}} + \ket{\overrightarrow{(6, 7)}} \right].
            \end{equation}
            The coin operator in this case is represented by
            \begin{equation}\label{coin_56}
                C = C_0 \bigoplus  C_1 \bigoplus C_2 \bigoplus C_3 \bigoplus C_4 \bigoplus -C_5 \bigoplus -C_6 \bigoplus C_7.
            \end{equation}
             \autoref{P2_B3_56_fidelity} shows that the fidelity is more than $0.8$ at time steps $t=57, 99, \dots $. The result is similar for the cases when $ s= 2$, $r = 5$; $ s= 2$, $r = 7$; $ s= 3$, $r = 4$; $ s= 3$ , $r = 6$ ; and $ s= 4$, $r = 7$.
            \item 
            \textbf{The sender and receivers are placed on the wings of the butterfly with the maximum distance, and they are in the same partite set.}

            For $s = 4$ and $r = 6$, we have
            \begin{equation}
            \ket{\psi(s)} = \frac{1}{\sqrt{2}} \left[\ket{\overrightarrow{(4, 0)}} + \ket{\overrightarrow{(4, 5)}} \right]; \text{and}~ \ket{\psi(r)} = \frac{1}{\sqrt{2}} \left[\ket{\overrightarrow{(6, 0)}} + \ket{\overrightarrow{(6, 7)}} \right].
            \end{equation}
            The coin operator is 
            \begin{equation}
            C = C_0 \bigoplus  C_1 \bigoplus C_2 \bigoplus C_3 \bigoplus -C_4 \bigoplus C_5 \bigoplus -C_6 \bigoplus C_7.
            \end{equation}

            As depicted in \autoref{P2_B3_46_fidelity}, state transfer fidelity is achieved more than $0.8$ at time steps $ t = 48, 162, \dots$. The similar behavior is observed for the cases when $ s= 2$, $r = 6$; $s= 2$, $r = 4$; $s= 3$, $r = 5$; $s= 3$, $r = 7$, and $s= 5$, $r = 7$. 
            \item 
            \textbf{One of the sender and receiver is placed on the body, and the other is placed on the wing of the butterfly, and they are in a different partite set.} 
            
            Here, assume that $s = 0$ and $r = 2$. 
            \begin{equation}
            \begin{split}
            \ket{\psi(s)} = &\frac{1}{\sqrt{4}} \left[\ket{\overrightarrow{(0, 1)}} + \ket{\overrightarrow{(0, 2)}} + \ket{\overrightarrow{(0, 4)}} + \ket{\overrightarrow{(0, 7)}}\right]; \\
            \ket{\psi(r)} = & \frac{1}{\sqrt{4}} \left[\ket{\overrightarrow{(2, 0)}} + \ket{\overrightarrow{(2, 3)}} \right].
            \end{split}
            \end{equation}
            The coin operator is given by
            \begin{equation}
                C = -C_0 \bigoplus  C_1 \bigoplus -C_2 \bigoplus C_3 \bigoplus C_4 \bigoplus C_5 \bigoplus C_6 \bigoplus C_7.
            \end{equation}
            In this case, fidelity is less than $0.5$ always, as depicted in Figure \ref{P2_B3_02_fidelity}. The maximum fidelity of state transfer is $0.4183$, which is at the time step $t = 175$.
        \end{enumerate}
        Now we calculate the average fidelity for different locations of the sender and the receiver on the butterfly graph $B_3$ generated by $P_2$. The average fidelity values in different cases are collected in \autoref{table_B_3_P_2}. The second-best fidelity is achieved at the maximum distance between the sender and receiver.

        \begin{table}[hbt!]
        \centering
        \small
        \begin{tabular}{|c|c|c|}
        \hline
        \textbf{Location of Sender} & \textbf{Receiver Node} & \textbf{Average Fidelity} \\
        \hline
         0 & 1 & 0.1698 \\
         \hline
         0 & 2 & 0.0406 \\
        \hline
         \textbf{5} & \textbf{6} & \textbf{0.0928} \\
        \hline
         4 & 6 & 0.0916 \\
        \hline
        \end{tabular}
        \caption{Average fidelity for different sender and receiver locations up to $200$ time-steps on the butterfly graph $B_{3}$ generated by path graph $P_{2}$.}
        \label{table_B_3_P_2}
        \end{table}
        
        \subsection{State transfer on the butterfly graphs generated by $P_3$}
        
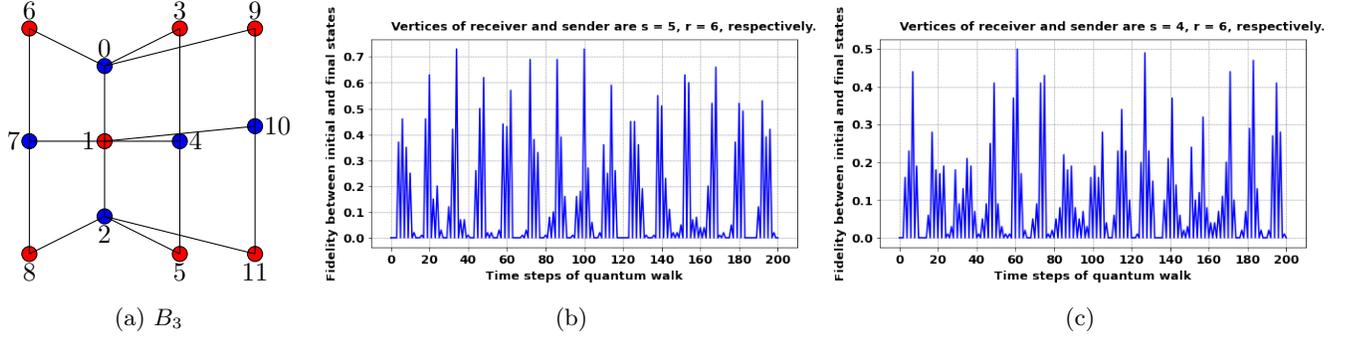
\begin{figure}
            \begin{subfigure}{.25\textwidth}
            \centering
            \begin{tikzpicture}
            \draw [fill = blue] (0, 1) circle [radius = 1mm];
            \node [above] at (0, 1) {${0}$};
            \draw [fill = red] (0, 0) circle [radius = 1mm];
            \node [left] at (0, 0) {${1}$};
            \draw [fill = blue] (0, -1) circle [radius = 1mm];
            \node [below] at (0, -1) {${2}$};
            \draw [fill = red] (1, 1.5) circle [radius = 1mm];
            \node [above] at (1, 1.5) {${3}$};
            \draw [fill = blue] (1, 0) circle [radius = 1mm];
            \node [right] at (1, 0) {${4}$};
            \draw [fill = red] (1, -1.5) circle [radius = 1mm];
            \node [below] at (1, -1.5) {${5}$};
            \draw [fill = red] (-1, 1.5) circle [radius = 1mm];
            \node [above] at (-1, 1.5) {${6}$};
            \draw [fill = blue] (-1, 0) circle [radius = 1mm];
            \node [left] at (-1, 0) {${7}$};
            \draw [fill = red] (-1, -1.5) circle [radius = 1mm];
            \node [below] at (-1, -1.5) {${8}$};
            \draw [fill = red] (2, 1.5) circle [radius = 1mm];
            \node [above] at (2, 1.5) {${9}$};
            \draw [fill = blue] (2, 0.2) circle [radius = 1mm];
            \node [right] at (2, 0.2) {${10}$};
            \draw [fill = red] (2, -1.5) circle [radius = 1mm];
            \node [below] at (2, -1.5) {${11}$};
            \draw (0, 1) -- (0, 0);
            \draw (0, 1) -- (1, 1.5);
            \draw (0, 0) -- (0, -1);
            \draw (0, 0) -- (1, 0);
            \draw (1, 1.5) -- (1,0);
            \draw (0, -1) -- (1, -1.5);
            \draw (1, 0) -- (1, -1.5);
            \draw (0,1) -- (-1, 1.5);
            \draw (0,0) -- (-1, 0);
            \draw (0, -1) -- (-1 , -1.5);
            \draw (-1 , 1.5) -- (-1, 0);
            \draw (-1 , 0) -- (-1 , -1.5);
            \draw (0, 1) -- (2, 1.5);
            \draw (0, 0) -- (2, 0.2);
            \draw (0 , -1) -- (2, -1.5);
            \draw (2, 1.5) -- (2, 0.2);
            \draw (2, 0.2) -- (2, -1.5);
            \end{tikzpicture}
            \caption{$B_{3}$}
            \label{B3_by_P3}
            \end{subfigure}
            \begin{subfigure}{0.38\textwidth}
            \includegraphics[width=\textwidth]{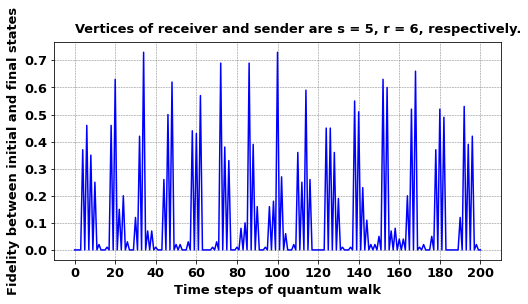}
                \caption{}
                \label{P3_B3_56_fidelity}
            \end{subfigure}
            \begin{subfigure}{0.38\textwidth}
            \includegraphics[width=\textwidth]{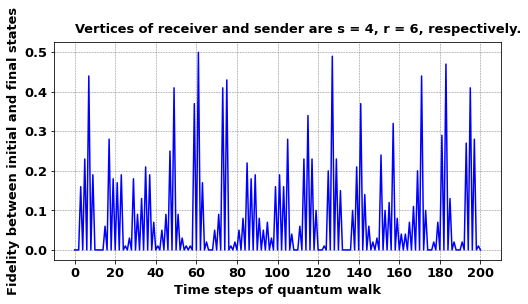}
                \caption{}
                \label{P3_B3_46_fidelity}
            \end{subfigure}
            \caption{State transfer on the butterfly graph $B_3$ formed by seed graph $P_{3}$.}
            \label{State_transfer_$P_3$}
        \end{figure}

        In this subsection, the seed graph $B_0 = P_3$ produces a family of butterfly graphs following Procedure  \ref{Butterfly_Procedure}. We consider only the butterfly graph $B_{3}$,  which can be constructed by adding three wings to the seed graph $B_{0}$. The graph $B_{3}$ is depicted in \autoref{B3_by_P3}.
         The quantum states of the sender and receiver can be found using the equations \eqref{sender_state} and \eqref{receiver_state}. The sender and receiver are located at different vertices of the network, and based on their positions and for better communication, the following cases are considered. 

        \begin{enumerate}
            \item 
            \textbf{The sender and receiver are placed on the wings of the butterfly with the maximum distance and are in the same partite set.} 
            
            Let the sender and the receiver be placed at $s = 5$ and $r = 6$, respectively. The similar results are obtained for the cases $ s= 6$, $r = 11$; $ s= 3 $, $r = 8$, and $ s= 8$, $r = 9$.
            \begin{equation}
            \ket{\psi(s)} = \frac{1}{\sqrt{2}} \left[\ket{\overrightarrow{(5, 2)}} + \ket{\overrightarrow{(5, 4)}} \right]; \text{and}~ \ket{\psi(r)} = \frac{1}{\sqrt{2}} \left[\ket{\overrightarrow{(6, 0)}} + \ket{\overrightarrow{(6, 7)}} \right].
            \end{equation}
            Following equation \eqref{coin}, the coin operator is given by
            \begin{equation}
                C = C_0 \bigoplus  C_1 \bigoplus C_2 \bigoplus C_3 \bigoplus C_4 \bigoplus -C_5 \bigoplus -C_6 \bigoplus C_7 \bigoplus C_8.
            \end{equation}
            From \autoref{B3_by_P3}, we notice that $d(0) = d(2) =4$, d(1) = 5. Also, the degree of all other vertices is $2$. The coin operator for nodes of degree four is given by equation \eqref{Coin_4}. The coin operators for the vertices with degree $5$ is,
            \begin{equation}
            C_{1} =\begin{bmatrix} \label{Coin_5}
                -0.6 & 0.4 & 0.4 & 0.4 & 0.4\\
                 0.4 & -0.6 & 0.4 & 0.4 & 0.4\\
                 0.4 & 0.4 & -0.6 & 0.4 & 0.4\\
                 0.4 & 0.4 & 0.4 & -0.6 & 0.4\\
                 0.4 & 0.4 & 0.4 & 0.4 & -0.6
            \end{bmatrix};
            \end{equation}
            which is the Grover's coin operator of dimension five.
            Similarly, the coin operators with degrees two and three can be calculated using equations \eqref{Coin_2} and \eqref{Coin_3}, respectively. We analyze the state transfer up to $200$ time steps and observe that the maximum fidelity of state transfer is $0.73$, which is at the time steps $t = 34$ and $t = 100$ as shown in \autoref{P3_B3_56_fidelity}. 
            \item 
            \textbf{The sender and receiver are placed on the wings of the butterfly with the maximum distance, and they are in different partite set.} 
            
            Consider $s = 4$ and $r = 6$. 
            Hence, the states of the sender and receiver are defined as,
            \begin{equation}
            \ket{\psi(s)} = \frac{1}{\sqrt{3}} \left[\ket{\overrightarrow{(4, 1)}} + \ket{\overrightarrow{(4, 3)}} + \ket{\overrightarrow{(4, 5)}} \right]; \text{and}~ \ket{\psi(r)} = \frac{1}{\sqrt{2}} \left[\ket{\overrightarrow{(6, 0)}} + \ket{\overrightarrow{(6, 7)}} \right].
            \end{equation}
            In addition, the coin operator is defined as,      \begin{equation}
                C = C_0 \bigoplus  C_1 \bigoplus C_2 \bigoplus C_3 \bigoplus -C_4 \bigoplus C_5 \bigoplus -C_6 \bigoplus C_7 \bigoplus C_8.
            \end{equation}
        \end{enumerate}
        The fidelity of the state transfer for this case is shown in \autoref{P3_B3_46_fidelity}. The maximum fidelity of state transfer occurs at time step $t= 61$, which is $0.5$. Also, we calculate the average fidelity between the sender and the receiver, placing them in different locations in the graph. The average fidelity values are collected in \autoref{Table_B_3_P_3}. Here, we interestingly notice that the maximum average fidelity occurs between the vertices $5$ and $6$. The distance between $5$ and $6$ is $4$, which is the maximum distance between a pair of vertices in the graph. Similarly the maximum fidelity is also observed when $ s= 6$, $r = 11$; $ s= 3 $, $r = 8$, and $ s= 8$, $r = 9$. Note that, in all the cases, the distance between the sender and the receiver is also $4$.
       
        \begin{table}[hbt!]
        \centering
        \small
        \begin{tabular}{|c|c|c|}
        \hline
        \textbf{Location of Sender} & \textbf{Receiver Node} & \textbf{Average Fidelity} \\
        \hline
         0 & 2 & 0.0992 \\
        \hline
         0 & 3 & 0.05775 \\
        \hline
         0 & 4 & 0.05465 \\
        \hline
         4 & 6 & 0.07215 \\
        \hline
         \textbf{5} & \textbf{6} & \textbf{0.1087} \\
        \hline
        \end{tabular}
        \caption{Average fidelity for different sender and receiver locations up to $200$ time-steps on the butterfly graph $B_{3}$ generated by path graph $P_{3}$. The maximum fidelity is observed between $5$ and $6$.}
        \label{Table_B_3_P_3}
        \end{table}

    \section{Quantum noise in higher dimensions}

    The dynamics of open quantum systems are described using quantum channels, which are completely positive trace-preserving (CPTP) maps. These channels can be expressed in terms of Kraus operators ${K_i}$, which satisfy the completeness relation    \begin{equation}\label{Kraus_operator_general}
    \sum_i K_i^\dagger K_i = I,
    \end{equation}
    where $I$ denotes the identity operator. If a quantum state $\rho$ evolves through a channel characterized by ${K_i}$, the output state is given by the operator-sum representation    \begin{equation}\label{completeness_condition}
    \rho' = \sum_{i} K_{i} \rho K_{i}^\dagger.
    \end{equation}
    
    To construct Kraus operators in arbitrary dimension, we employ the framework of Weyl operators \cite{bertlmann2008bloch, weyl1927quantenmechanik}, which generalize the Pauli operators to higher-dimensional systems \cite{dutta2023qudit, basile2024weyl}. The Weyl operators of order $d$ are   \begin{equation}\label{Weyl_operator}
    U_{u,v}= \sum_{k=0}^{d-1} e^{(\frac{2\pi i}{d})ku} \ket{k}\bra{(k+v) \bmod d}, ~\text{where}~ 0 \leq u, v \leq (d-1).
    \end{equation}
    We can prove that $U_{u,v}$ is unitary for all $v$ and $u$, that is $U_{u,v}^\dagger U_{u,v} =  U_{u,v} U_{u,v}^\dagger = I_d$. In particular, for $u=v=0$, we obtain the identity operator $U_{0,0} = I_d$. 
    
    In this work, we focus on two types of paradigmatic models of quantum noise: non-Markovian unital noises such as the RTN and OUN, as well as non-unital non-Markovian ADN. The RTN is characterized by an oscillatory and damped behavior in its non-Markovian regime, whereas the OUN exhibits a monotonic power-law decay without oscillations \cite{utagi2020temporal}. ADN models the physical process of energy loss from a quantum system to its surrounding environment. In simple terms, it describes how an excited quantum state gradually relaxes to a lower energy state due to dissipation into the environment. These noise models were originally formulated for qubit systems. However, for DTQWs on graphs with arbitrary numbers of vertices and edges, it becomes necessary to extend these noise processes to higher-dimensional quantum systems. To achieve this, we construct generalized quantum channels using Weyl operators $U_{u,v}$, whose dimensionality corresponds to the order of the underlying graphs under investigation.
    
    The non-Markovian RTN \cite{kumar2018non, banerjee2017non} can be represented by the Kraus operators 
    \begin{equation}\label{RTN}
    \begin{split}
    & K_1(t) = \sqrt{\frac{1+ \Lambda(t)}{2}}~ U_{0,0}, ~\text{and}~ K_2(t) = \sqrt{\frac{1-\Lambda(t)}{2}}~ U_{1,0}, ~\text{where}\\
    & \Lambda(t) = e^{-\gamma t} \left[ \cos\left(\nu \gamma t \right) + \frac{\sin\left(\nu \gamma t \right)}{\nu} \right], ~\text{and} ~ \nu = {\sqrt{\left( \frac{2a}{\gamma} \right)^2 - 1}}.
    \end{split}
    \end{equation}
    Here, $\nu$ denotes the characteristic frequency of the harmonic oscillator, while $\Lambda(t)$ describes the damped harmonic modulation governing the system-environment interaction. The dynamics exhibit non-Markovian behavior when the ratio $\frac{a}{\gamma} > 0.5$, indicating a regime of information backflow from the environment to the system. For fixed parameters $a$ and $\gamma$, the resulting channel is explicitly time-dependent, with its evolution determined by the parameter $t$.
    
    The non-Markovian OUN also induces pronounced memory effects in the system dynamics \cite{uhlenbeck1930theory, kumar2018non, banerjee2017non}. For quantum states defined in arbitrary dimensions, the corresponding channel can be expressed by the Kraus operators
    \begin{equation}\label{OUN}
    \begin{split}
    & K_1(t) = \sqrt{\frac{1+ P(t)}{2}}~ U_{0,0}, ~\text{and}~ K_2(t) = \sqrt{\frac{1- P(t)}{2}}~ U_{1,0}, ~\text{where}\\
    & P(t) = e^{-\frac{\lambda}{2} \left(t + \frac{1}{\gamma} \left(e^{-\gamma t} - 1\right)\right)}.
    \end{split}
    \end{equation}
    Here, $\gamma$ characterizes the noise bandwidth, while $\lambda$ denotes the relaxation time of the environment. When both $\gamma$ and $\lambda$ are treated as constants, the dynamics of the channel become explicitly dependent only on the parameter $t$. In our analysis, we fix the parameters to $\lambda = 1$ and $\gamma = 0.05$ for all the graph structures under consideration. 

    The non-Markovian ADN acting on an arbitrary $2m$ dimensional quantum system can be expressed through the following Kraus operators \cite{dutta2023qudit},
    \begin{equation}
    \begin{split}
        K_1 & = |0 \rangle \langle 0| + \sqrt{1-\lambda(t)} \sum_{j = 1}^{2m - 1} |j \rangle \langle  j| \\
        K_j & = \sqrt{\lambda(t)} |0 \rangle \langle j| ~\text{for}~ 1 \leq j \leq 2m - 1.
    \end{split}
    \end{equation}
    Here, $\lambda(t) = 1 - e^{-gt}\left[ \frac{g}{l}\sinh\left(\frac{lt}{2}\right) + \cosh\left(\frac{lt}{2}\right) \right]^2$ with $l = \sqrt{g^2 - 2 \gamma g}$. Also, $g$ represents the spectral width of the system-environment coupling, while $\gamma$ represents the spontaneous emission rate. In this article, throughout our analysis of graph structures under noise, the parameters are kept fixed at $g = 0.001$ and $\gamma = 5$. Also, $\ket{0}$ and $\ket{j}: j = 1, 2, \dots (2m - 1)$ are the $2m$ dimensional state vectors in the computational basis of $\mathcal{H}^{(2m)}$, where $m$ is the number of undirected edges in a graph $G$.

    \section{Quantum state transfer under noise}

        To understand the effect of quantum noise on quantum state transfer we consider two graphs, which are the butterfly graph $B_3$ generated by $P_2$ and the butterfly graph $B_3$ generated by $P_3$. These graphs are depicted in \autoref{B3_by_P2} and \autoref{B3_by_P3}, respectively. Recall that in \autoref{Discrete-time_quantum_walk_on_the_butterfly_graphs} we have already discussed their state transfer properties in the absence of quantum noise.

        \subsection{Effect of noise on state transfer on butterfly graph $B_3$ generated by $P_{2}$}
            
            In subsection 4.1.4, we considered four combinations of the sender and the receiver on the graph. Here, we consider the case when the state transfer fidelity is maximal in the absence of noise. For this, let the sender and receiver be located at the vertices $s = 5$ and $r=6$. Now, we calculate $\ket{\psi_t} = U_{\overrightarrow{5, 6}}^t \ket{\psi(s)}$, where $U_{\overrightarrow{5, 6}} = S \times C$. We can calculate $S$ from equation \eqref{shift} and $C$ is mentioned in equation \eqref{coin_56}. Note that, $\ket{\psi_t}$ is a pure state for all $t$. Applying equation \eqref{Fidelity_vectors}, we can calculate the fidelity between $\ket{\psi(r)}$ and $\ket{\psi_t}$ for different values of $t$. 
            
\begin{figure}
                \begin{subfigure}{0.33\textwidth}   \includegraphics[width=\textwidth]{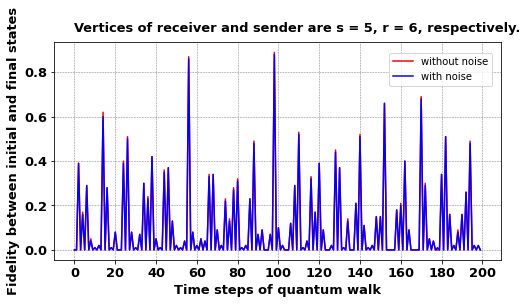}
                \caption{}
                \label{P2_B3_RTN}
                \end{subfigure}
                \begin{subfigure}{0.33\textwidth}
                \includegraphics[width=\textwidth]{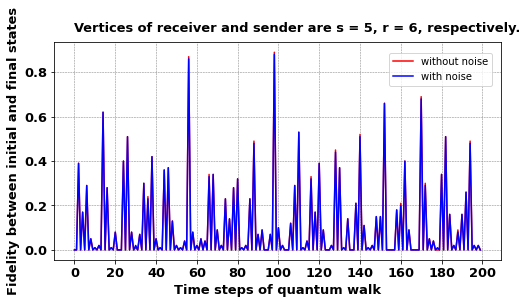}
                \caption{}
                \label{P2_B3_OUN}
                \end{subfigure}
                \begin{subfigure}{0.33\textwidth}
                \includegraphics[width=\textwidth]{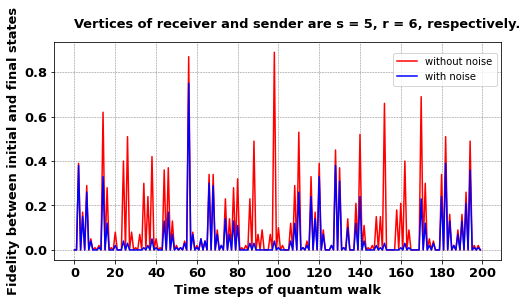}
                \caption{}
                \label{P2_B3_NMAD}
                \end{subfigure}
                \caption{State-transfer on the butterfly graph $B_{3}$ generated by path graph $P_{2}$, in the absence of noise and under non-Markovian RTN (\ref{P2_B3_RTN}), modified non-Markovian OUN (\ref{P2_B3_OUN}) and NMAD noise (\ref{P2_B3_NMAD}). The RTN parameters are set to $a = 0.1$, $\gamma = 0.01$, while the OUN parameters are $\lambda = 1$, $\gamma = 0.05$. The channel parameters for NMAD noise are $\gamma = 5$ and $g = .001$.}
                \label{P2_B3_state_transfer}
            \end{figure}

            To apply quantum noise on a quantum walk, we send the state via a quantum channel. Therefore, if at time $t$, the state of the walker be $\ket{\psi_t}$, the state after applying quantum noise becomes $\rho_t^{'}$, following equation \eqref{completeness_condition}. Let the channel is given by the Kraus operators $K_i$, then 
            \begin{equation}
                \rho_t^{'} = \sum K_i \ket{\psi_t}\bra{\psi_t}K_i^{\dagger}.
            \end{equation}
            If $K_i$ are time-dependent, we consider the time $t$ to calculate the expression of the Kraus operator.
            \autoref{P2_B3_state_transfer} shows the fidelity of state transfer between $s = 5$ and $r = 6$ with and without noise. In the case of non-Markovian RTN and OUN noises, the oscillatory behavior of the fidelity is largely preserved. The peaks in the noisy case closely follow those of the noiseless evolution, indicating that the state transfer process remains relatively robust. In contrast, under NMAD noise, the fidelity peaks are noticeably suppressed compared to the noiseless case. This reduction occurs because amplitude damping represents a dissipative interaction with the environment, leading to a partial loss of quantum coherence. However, due to the non-Markovian nature of the noise, small revivals in fidelity are still observed, reflecting the temporary backflow of information from the environment to the system. Hence, the quantum state transfer on the butterfly graph $B_3$ constructed from $P_2$ is more strongly affected by non-Markovian amplitude damping noise, while non-Markovian RTN noise has a comparatively weaker impact, allowing the transfer dynamics to remain closer to the ideal noiseless behavior. This is due to the fact that non-Markovian amplitude damping (NMAD) noise exhibits both dissipative and dephasing-like effects, whereas unital noise channels exhibit only dephasing behavior.
            
\begin{figure}
                \begin{subfigure}{0.33\textwidth}   \includegraphics[width=\textwidth]{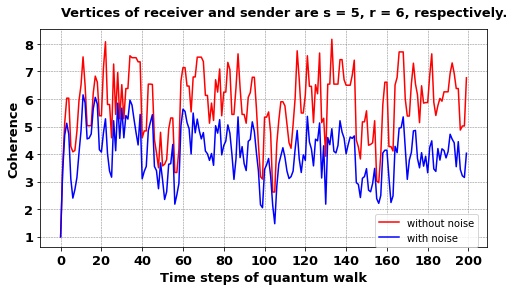}
                \caption{}
                \label{P2_B3_Coherence_RTN}
                \end{subfigure}
                \begin{subfigure}{0.33\textwidth}
                \includegraphics[width=\textwidth]{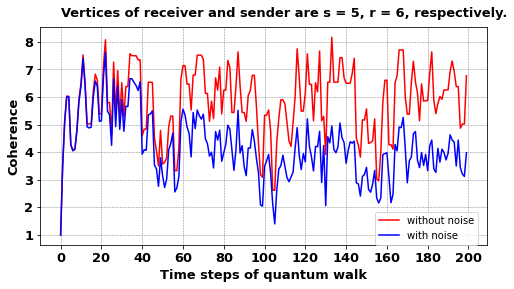}
                \caption{}
                \label{P2_B3_Coherence_OUN}
                \end{subfigure}
                \begin{subfigure}{0.33\textwidth}
                \includegraphics[width=\textwidth]{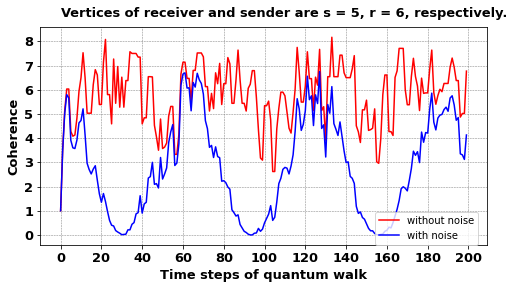}
                \caption{}
                \label{P2_B3_Coherence_ADC}
                \end{subfigure}
                \caption{Coherence of a quantum walker for a butterfly graph $B_{3}$ generated by path graph $P_{2}$, in the absence of noise and under non-Markovian RTN (\ref{P2_B3_Coherence_RTN}), modified non-Markovian OUN (\ref{P2_B3_Coherence_OUN}) and non-Markovian amplitude damping noise (\ref{P2_B3_Coherence_ADC}). The channel parameters are the same as before.}
                \label{P2_B3_Coherence}
            \end{figure}

            Recall that after $t$ steps of quantum walk, the state of the walker is $\ket{\psi_t}$. After applying quantum noise, we get the density matrix $\rho_t'$. We calculate the coherence of the quantum states $\ket{\psi_t}$ and $\rho_t'$.  We observe $l_1$ norm of coherence of a quantum walker, in the absence of noise and under non-Markovian unital and non-unital noises, respectively, as shown in \autoref{P2_B3_Coherence}. In the case of RTN, coherence leads to irregular fluctuations. On the other hand, under OUN, coherence decays smoothly and then stabilizes around a finite value with relatively mild fluctuations, indicating a more controlled loss of quantum superposition due to finite environmental memory. In the presence of non-unital ADC, coherence shows smooth oscillatory revivals, indicating a periodic backflow of information from the environment that temporarily restores quantum superposition. The impact of non-unital noise is higher than that of unital noise.

       \subsection{Effect of noise on state transfer on butterfly graph $B_3$ generated by the $P_{3}$}
       
\begin{figure}
          \begin{subfigure}{0.33\textwidth}
            \includegraphics[width=\textwidth]{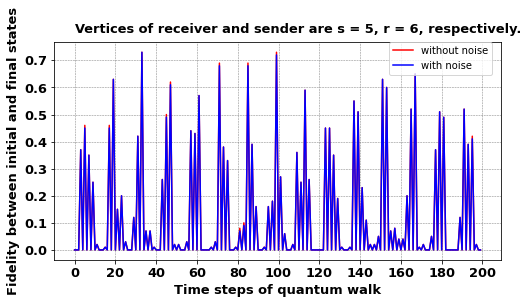}
                \caption{}
                \label{P3_B3_RTN}
          \end{subfigure}
          \begin{subfigure}{0.33\textwidth}
            \includegraphics[width=\textwidth]{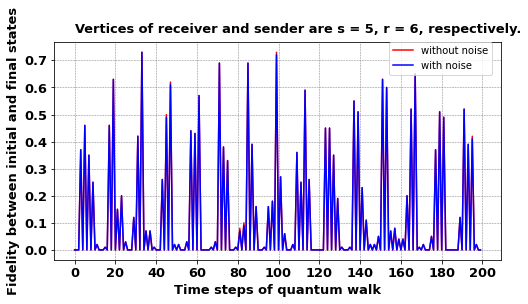}
                \caption{}
                \label{P3_B3_OUN}
          \end{subfigure}
          \begin{subfigure}{0.33\textwidth}
            \includegraphics[width=\textwidth]{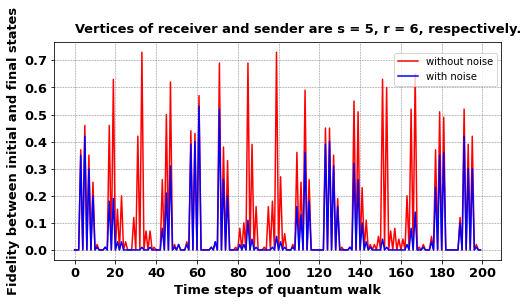}
                \caption{}
                \label{P3_B3_NMAD}
          \end{subfigure}
        \caption{State transfer on the the butterfly graph $B_{3}$ generated by path graph $P_{3}$, in the absence of noise and under non-Markovian RTN (\ref{P3_B3_RTN}), non-Markovian OUN noise (\ref{P3_B3_OUN}) and  non-Markovian amplitude damping noise (\ref{P3_B3_NMAD}). The channel parameters are the same as before.}
           \label{P3_B3_state_transfer}
       \end{figure}

        In this subsection, we observe the effect of noise on the butterfly graph $B_{3}$ under different non-Markovian noises, including unital and non-unital noises such as RTN, OUN and ADN generated by the seed graph $B_0 = P_3$. Here, we also consider $s = 5$ and $r = 6$ as the location of the sender and the receiver. In \autoref{P3_B3_state_transfer}, we plot the fidelity in noisy and noiseless scenarios. From this figure, we observe that the fidelity remains low for most time steps with irregular peaks, while it is high at a few steps. Also, the oscillatory behavior of the fidelity is largely preserved, and several peaks retain amplitudes comparable to the noiseless case. This suggests that RTN and OUN noises mainly introduce fluctuations in the evolution rather than strong dissipation, allowing the system to maintain higher coherence. In contrast, in the case of NMAD noise, the fidelity peaks are significantly suppressed compared to the noiseless evolution, indicating a stronger degradation of quantum coherence due to dissipative interaction with the environment. However, small revivals occur because of the memory effects of the non-Markovian dynamics. Hence, quantum state transfer on the butterfly graph $B_3$ is more strongly affected by NMAD noise than by non-Markovian RTN and OUN noises. We also observe that using the path graph $P_{3}$ as a seed yields poorer quantum state transfer between the sender and receiver than using the path graph $P_{2}$ as a seed.
    
    \section{Conclusion}
    
    
    
    In Procedure \ref{Butterfly_Procedure}, we discussed the process for generating the family of Butterfly graphs, which allowed us to construct a scalable network. In this investigation, we established that the butterfly graphs generated from path graphs support quantum state transfer under discrete-time quantum walks, thereby extending the class of graph structures enabling efficient quantum transport.

    Furthermore, we have systematically investigated the robustness of quantum state transfer in these butterfly graphs generated by the path graph under the influence of different non-Markovian noise models. In particular, we analyzed the effects of unital noise channels such as non-Markovian RTN and non-Markovian OUN, along with the non-unital non-Markovian ADN and compared with the corresponding noiseless evolution through the fidelity dynamics. Under unital noises, there is no significant effect of noise on state transfer on the butterfly graphs with time steps. While under unital noise, there is a larger impact.

    In addition to fidelity, we have explored the behavior of quantum coherence, which serves as a fundamental resource for quantum information processing. In the presence of noise, coherence generally decays over time due to environmental interactions. For unital noise channels, coherence shows non-monotonic behavior with partial revivals, consistent with their non-Markovian character. The non-unital noise has a higher impact. Overall, the combined analysis of fidelity and coherence offers deeper insight into quantum state transport in noisy environments. Our results show that both graph structure and environmental noise strongly affect the efficiency of quantum state transfer, providing useful guidelines for designing robust quantum communication systems.

    \section*{Funding}
    This work is supported by the SERB funded project entitled “Transmission of quantum information using perfect state transfer” (Grant no. CRG/2021/001834).
    

\end{document}